\documentclass[aps,12pt,preprintnumbers,nofootinbib,superscriptaddress]{revtex4}
\usepackage{graphicx}
\usepackage{amssymb,amsmath}
\def\Dslash{D\hskip-0.65em /}

\begin{document}

\title{\Large{\bf Higher Derivative Operators in the Noncommutative Schwinger Model } }

\author{Kaniba Mady Keita}
\author{Feng Wu}
\email[Electronic address: ]{fengwu@ncu.edu.cn}
\affiliation{%
Department of Physics, Nanchang University,
330031, China}
\author{Ming Zhong}
\email[Electronic address: ]{zhongm@nudt.edu.cn}
\affiliation{%
Department of Physics, National
University of Defense Technology, Hunan 410073, China}

\date{\today}

\begin{abstract}
A study of the noncommutative Schwinger model is presented. It is shown that the Schwinger mass is not modified by the noncommutativity of spacetime till the first nontrivial order in the noncommutative parameter. Instead, a higher derivative kinetic term is dynamically generated by the lowest-order vacuum polarization diagrams. We argue that in the framework of the Seiberg-Witten map the feature of non-unitarity for a field theory with space-time noncommutativity is characterized by the presence of higher derivative kinetic terms. The $\theta$-expanded version of a unitary theory will not generate the lowest-order higher derivative quadratic terms. 
\end{abstract}
\maketitle 
\newpage

\date{\today}

\section{Introduction}
Although no experimental evidence has yet been detected, it is possible that coordinates of spacetime are noncommutative \cite{snyder}. One theoretical support comes from the fact that field theories constructed on a Moyal space can be viewed as low-energy effective theories from open string theory with a constant NS-NS $B$ field \cite{s-w}. In the context of field theories space-time coordinates satisfy   
\begin{equation}
[x^{\mu}, x^{\nu}]=i \theta^{\mu\nu} \label{noncomu}
\end{equation}
where $\theta^{\mu\nu}$ is a real constant antisymmetric matrix. Energy-momentum conservation still holds from translational invariance in the Moyal space. Usually the commutation relations~(\ref{noncomu}) spoil Lorentz invariance. However, it is clear that both Lorentz symmetry and translation symmetry are preserved in $(1+1)$ dimensional spacetime. 

It is shown in Ref.\cite{s-w} that a noncommutative gauge theory should be gauge equivalent to an ordinary counterpart defined on a commutative spacetime. The two equivalent descriptions are related to each other by the Seiberg-Witten map. After that noncommutative field theories have been an extensively studied subject and many properties have been learned. In particular, the unitarity of scalar field theories in noncommutative spacetime was investigated~\cite{gomis} in the approach of covariant perturbation theory and it was shown that the uncertainty relation between temporal and spatial coordinates would lead to non-unitarity of quantum field theories. In other words, noncommutative field theories with $\theta^{0i} \neq 0$ do not satisfy Cutkosky's cutting rules\footnote{Although it has been shown that~\cite{bahns,liao} unitarity is manifest in the approach of time-ordered perturbation theory, this formalism causes other serious flaws~\cite{ohl,reichenbach}. In this paper we restrict our discussion in the context of covariant perturbation theory.}. Therefore, noncommutative quantum field theories in two dimensional spacetime are not unitary. 

Unitarity was traditionally treated as a criterion to judge the fate of a quantum field theory. A theory that violates unitary contains as a part the states with negative norm. However, a modern point of view within the framework of effective field theories is that a field theory violating unitarity might still be sensible at a low energy as long as the ghost states are unstable so that one cannot have them as asymptotic states (for more details, see Ref.~\cite{antoniadis}).

The Schwinger model \cite{schwinger}, quantum electrodynamics of massless fermions in two dimensional spacetime, has been a subject of interest for a long time (see Ref.~\cite{lowenstein} for a review). Ever since Schwinger's pioneer work, it is known that in two dimensional spacetime the theory of a massless Dirac fermion $\psi$ is equivalent to the theory of a massless scalar field $\phi$. In particular, the current operator $\bar{\psi} \gamma_{\mu} \psi$ is equivalent to ${1\over \pi} \epsilon_{\mu\nu} \partial^{\nu} \phi$ and the chiral composites $\bar{\psi}(1\pm \gamma_{5} )\psi$ are $e^{\pm i \sqrt{4 \pi} \phi}$ up to a prescription dependent constant \cite{coleman}.
The emergence of the phenomena such as confinement of fermions, gauge boson mass generation, axial anomaly, and the nontrivial topological sectors has made this exactly solvable model a productive theoretical laboratory. Easier calculations in the Schwinger model than other gauge theories provides theorists a testing ground for new ideas.

The aim of this paper is to study the Schwinger model in a Moyal spacetime manifold. We show that, after the Seiberg-Witten map, up to the first nontrivial order in the noncommutative parameter, the counterpart of the noncommutative Schwinger model has only one additional term compared with the ordinary Schwinger model. This term is of the form $\theta^{\alpha\beta}F_{\beta\gamma}F^{\gamma\delta}F_{\delta\alpha}$. In other words, the fermion sector is not modified by the noncommutative structure of spacetime. It should be stressed that the above statement is only correct in two dimensional spacetime. As is well known, the fermion anomaly has many important applications in physics. Since the fermion sector is the same as the ordinary Schwinger model, the ABJ anomaly is left unchanged up to the first nontrivial order in the noncommutative parameter . Moreover, we will show perturbatively that the gauge boson mass is not modified by the self-energy diagram constructed from the three-photon vertex. 

Note that the origin of the gauge boson self-interaction terms in an ordinary non-Abelian gauge theory are due to the noncommutativity of the internal space. Here the photon self-interaction term is due to the noncommutative structure of the external spacetime. What are the physical effects under the presence of the photon self interaction term? The subject of this paper is to answer this question. Indeed, we will show by explicit calculations that a higher derivative operator of Lee-Wick type \cite{lee-wick} is generated dynamically. Once this is done, we argue that the noncommutativity between temporal and spatial coordinates provides a natural explanation of the presence of higher derivative operators in theories such as Lee-Wick QED \cite{lee-wick} and Lee-Wick standard model \cite{wise} (see Ref.~\cite{wu} for another prescription).

For early works on the noncommutative Schwinger model, see Ref.~\cite{grosse}.

The rest of the paper is organized as follows. In the next section we consider the $\theta$-expansion of the noncommutative Schwinger model by first studying $(n+1)$ dimensional noncommutative QED. In this way one can see manifestly which features are dimensionality-dependent. Global and local symmetries are discussed. Most of the remarks in this section are well-known. We then proceed with calculating the one-loop vacuum polarization diagrams in section III. We show that because of the noncommutativity of spacetime a higher derivative operator is generated and the photon mass is not modified by the structure of spacetime. In section IV we discuss the relation between the emergence of higher derivative operators and the unitarity of a noncommutative field theory. We conclude our investigations in the last section.

\section{From $(n+1)$ noncommutative QED to noncommutative Schwinger model }
We start with quantum electrodynamics of massless fermions in a noncommutative $\mathbb{R}^{1,n} $ space characterized by Eqn.~(\ref{noncomu}). Its action is given by 
\begin{equation}
S=\int d^{n+1} x \left( -{1\over 4} \hat{F}_{\mu\nu} \star \hat{F}^{\mu\nu} + \hat{\bar{\psi}} \star i \hat{\Dslash}\hat{\psi} 
 \right) \label{action1}
\end{equation}
where the field strength of the gauge connection $\hat{A}_{\mu}$ and the covariant derivative are defined as
\begin{eqnarray}
\hat{F}_{\mu\nu} &=& \partial_{\mu} \hat{A}_{\nu} - \partial_{\nu} \hat{A}_{\mu} - i g [ \hat{A}_{\mu} , \hat{A}_{\nu} ]_{\star}, \\
\hat{D}_{\mu} \hat{\psi} &=& \partial_{\mu} \psi -ig \hat{A}_{\mu} \star \hat{\psi}.
\end{eqnarray}
Here $[A,B]_{\star}$ denotes Moyal bracket:
\begin{equation}
[A,B]_{\star} \equiv A \star B - B \star A .
\end{equation}
The $\star$-product of two fields $\phi_{1}(x)$ and $\phi_{2}(x)$ is defined as:
\begin{equation}
\phi_{1}(x) \star \phi_{2}(x) = exp \left( i {\theta^{\mu\nu} \over 2} {\partial \over \partial \xi^{\mu}}{\partial \over \partial \eta^{\nu}} \right) \phi_{1}(x+\xi) \phi_{2}(x+\eta) \vert_{\xi=\eta=0}.
\end{equation}

The coupling constant $g$ has mass dimension $({3-n \over 2})$. Therefore in four dimensional spacetime it is dimensionless while in two dimensional spacetime it has dimension 1.

The action~(\ref{action1}) is invariant under the noncommutative $U(1)$ gauge transformations:
\begin{eqnarray}
\hat{\delta}_{\hat{\lambda}} \hat{A}_{\mu} &=& {1\over g} \partial_{\mu} \hat{\lambda} +i [ \hat{\lambda} , \hat{A}_{\mu}]_{\star},\\
\hat{\delta}_{\hat{\lambda}} \hat{\psi} &=& i \hat{\lambda} \star \hat{\psi},\\
\hat{\delta}_{\hat{\lambda}} \hat{\bar{\psi}} &=&-i \hat{\bar{\psi}} \star \hat{\lambda}.
\end{eqnarray}
To the lowest order in the noncommutative parameter, the Seiberg-Witten map gives \cite{s-w} 
\begin{eqnarray}
\hat{A}_{\mu}&=& A_{\mu} - {g\over 2} \theta^{\rho\sigma} A_{\rho} \left( \partial_{\sigma} A_{\mu} + F_{\sigma\mu} \right) +O(\theta^{2}),\\
\hat{\psi}&=& \psi -{g\over 2} \theta^{\rho\sigma} A_{\rho} \partial_{\sigma} \psi +O(\theta^{2}),\\
\hat{\lambda}&=& \lambda - {g\over 2} \theta^{\rho \sigma} A_{\rho} \partial_{\sigma} \lambda +O(\theta^{2}),
\end{eqnarray}
where $A_{\mu}$, $\psi$ and $\lambda$ are the ordinary gauge field, fermion field and gauge transformation parameter respectively.

Using these expressions, up to order $O(\theta)$ \cite{jurco}, we have
\begin{equation}
\begin{split}
S= \int d^{n+1} x & \left(  -{1\over 4} \left[ \left( 1+{g\over2} \theta^{\mu\rho}F_{\rho\mu} \right) F^{\alpha\beta}F_{\alpha\beta} +
2g \theta^{\mu\rho} {F_{\rho}}^{\nu} {F_{\mu}}^{\sigma} F_{\sigma\nu} \right] \right. \\
&\left. + \left(1+{g\over 4} \theta^{\mu\rho}F_{\rho\mu}\right) \bar{\psi} i \Dslash \psi -{g\over 2} \theta^{\alpha\beta} \bar{\psi} i \gamma^{\mu} F_{\mu\alpha} D_{\beta} \psi \right).\label{1staction}
\end{split}
\end{equation}
This expansion makes sense since the noncommutativity of spacetime, if exists, is small. The second and third terms in~(\ref{1staction}) are the photon self-interaction terms. Similar to Yang-Mills theories in commutative spacetime, the existence of gauge boson self-interaction terms cause the theory to be asymptotically free \cite{martin}. However, in noncommutative quantum electrodynamics it is the structure of spacetime causing the nonlinearity of the field strength in the gauge connection. 

A remarkable fact is that for $n > 1$ all the $\theta$-dependent terms in Eqn.~(\ref{1staction}) are Lorentz violating. Thus in four dimensional spacetime the action ~(\ref{1staction}) provides an interesting Lorentz-violating extension of QED. For further details on this topic, we refer to Ref.~\cite{carroll}. 

The canonical momenta conjugate to the fermion field $\psi$ and the gauge field $A_{\mu}$ are
\begin{equation}
\Pi_{\psi}\equiv {\partial \mathcal{L} \over \partial \dot{\psi}}=\left( 1+{g\over 4} \theta^{\mu\rho} F_{\rho\mu} \right) i \psi^{\dagger} -{g\over 2} \theta^{\alpha 0}\bar{\psi} i \gamma^{\mu} F_{\mu\alpha},\label{pi}
\end{equation}
and
\begin{equation}
\begin{split}
\Pi^{\mu}  \equiv {\partial \mathcal{L} \over \partial \dot{A}_{\mu}} =  & {g\over 4} \theta^{0 \mu } F^{\alpha\beta} F_{\alpha\beta} +\left( 1+ {g\over 2} \theta^{\mu\rho} F_{\rho\mu} \right)F^{ \mu 0}- g \left( \theta^{\lambda [ 0} F_{\lambda \sigma} F^{\sigma \mu ]} + \theta^{\lambda \rho} {F_{\rho}}^{\mu} {F_{\lambda}}^{0}\right) \\
& +{g\over 2} \theta^{\mu 0}\bar{\psi} i \Dslash \psi + {g\over 2} \theta^{[ 0 \beta} \bar{\psi} i \gamma^{\mu ] } D_{\beta} \psi.\label{piA}
\end{split}
\end{equation}
It is obvious to see from the above result that $\Pi^{0} = 0$, which means that $A_{0}$ does not propagate. This result is the same as the one in commutative quantum electrodynamics. The well-known physical meaning of this is that there are redundant modes in the Lagrangian. Eqn.~(\ref{pi}) and Eqn.~(\ref{piA}) show that in $(n+1)$ noncommutative spacetime the canonical momenta $\Pi_{\psi}$ and $\Pi^{\mu}$ depend on both the gauge connection field $A_{\mu}$ and the fermion field $\psi$. However, as is shown in the following, in two dimensional spacetime, commutative or not, $\Pi_{\psi}$ depends only on the fermion field and $\Pi^{\mu}$ depends only on the gauge connection field.    
 
The equations of motion for the Lagrangian in~(\ref{1staction}) are
\begin{equation}
\left(1+{g\over 4} \theta^{\mu\rho} F_{\rho\mu} \right) i \Dslash \psi - {g \over 2} \theta^{\alpha\beta} F_{\mu\alpha} i \gamma^{\mu} D_{\beta} \psi =0,
\end{equation}
and
\begin{equation}
\begin{split}
 \partial_{\xi}& \left( - F^{\xi \lambda} \left(1+ {g \over 2} \theta^{\mu\rho} F_{\rho \mu} \right) +{g \over 4} \theta^{ \xi \lambda} F^{\alpha\beta} F_{\alpha\beta} - g \left( \theta^{\mu [ \xi} F_{\mu \sigma} F^{\sigma \lambda ]} + \theta^{\mu\rho} {F_{\rho}}^{\lambda} {F_{\mu}}^{\xi} \right) \right. \\
&\left. -{g \over 2} \left(\theta^{\xi \lambda} \bar{\psi} i \Dslash \psi - \theta^{[ \xi \beta} \bar{\psi} i \gamma^{\lambda ]} D_{\beta} \psi \right)   \right) =  g \left( 1+ {g\over 4} \theta^{\mu\rho} F_{\rho\mu} \right) \bar{\psi} \gamma^{\lambda} \psi - {g\over 2} \theta^{\alpha \lambda} \bar{\psi} \gamma^{\mu} F_{\mu \alpha} \psi .
\end{split}
\end{equation}
For $n \neq 1$, the above equations are the Lorentz-breaking extensions of the Dirac equation and the inhomogeneous Maxwell equations in the presence of sources.

At this point, the dimensionality of spacetime is not specified. From now on, we will focus on two dimensional spacetime. Our conventions for the spacetime are
\begin{equation}
\theta^{\mu\nu}= \theta \epsilon^{\mu\nu}
\end{equation}
with $\epsilon^{01}=g^{00} = - g^{11} =1$. The dimension of the Lorentz invariant parameter $\theta$ is length-squared. Thus $\sqrt{\theta} $ is related to the noncommutativity length scale. In two dimensional spacetime, the action~(\ref{action1}) is the noncommutative Schwinger model. It is straightforward to show from Eqn.~(\ref{1staction}) that in this context the Lagrangian is 
\begin{equation}
\begin{split}
\mathcal{L} &= - {1\over 4} \left( F_{\alpha\beta} F^{\alpha\beta} + g \theta \epsilon^{\alpha \beta}F_{\beta\gamma} F^{\gamma\delta} F_{\delta\alpha} \right) + \bar{\psi} i \Dslash \psi \\
&=- {1\over 4} \left( 1+{g\over 2} \theta \epsilon^{\alpha\beta} F_{\alpha\beta} \right) F_{\mu\nu} F^{\mu\nu}  + \bar{\psi} i \Dslash \psi . \label{schwinger}
\end{split} 
\end{equation}
From the first line to the second line in Eqn.~(\ref{schwinger}), we used the relation $\epsilon^{\alpha\beta}\epsilon_{\beta\gamma}\epsilon^{\gamma\delta}\epsilon_{\delta\alpha}={1\over 2} ( \epsilon^{\alpha\beta} \epsilon_{\alpha\beta} )^2$, which is correct in two dimensional spacetime. 

Therefore, with spacetime noncommutativity, up to the first nontrivial order in $\theta$ we end up with a ``modified" Schwinger model. The difference between the Lagrangian~(\ref{schwinger}) and the Lagrangian of the ordinary Schwinger model is the three-photon interaction term. Besides the local $U(1)$ gauge symmetry, the Lagrangian~(\ref{schwinger}) is invariant under global $U(1) \otimes U(1)^{\prime}$ symmetries, where $U(1)$ and $U(1)^{\prime}$ are the charge and chirality symmetries of the massless fermion. 

Discrete symmetries of noncommutative quantum electrodynamics have been investigated in \cite{sheikh}. One can do the same analysis to the ``modified" Schwinger model defined by~(\ref{schwinger}) . It is not difficult to show that it is parity ($P$) invariant. However, both charge conjugation ($C$) and time reversal ($T$) are violated by the three-photon interaction term. A theory with a positive noncommutative parameter is related to the one with a negative noncommutative parameter by charge conjugation. In consequence, the theory is still $CPT$ invariant but $C$ and $CP$ violated. Note that since the noncommutative Schwinger model is manifestly Lorentz invariant, different from the four dimensional noncommutative quantum electrodynamics, $CPT$ invariance is not a surprising result.

One interesting result in Eqn.~(\ref{schwinger}) is that the fermion sector is not modified in $(1+1)$ spacetime at one loop order. This means that different from the gauge boson, a two dimensional fermion does not know whether the spacetime is commutative or not and will respond in the same way in either case. Since there is an exact mapping between the fermionic and bosonic theories in two dimensional spacetime, the same is true for a scalar. From this one can easily conclude that the ABJ anomaly is the same as the ordinary result till the first nontrivial order in $\theta$. This agrees with the analysis in Ref.~\cite{banerjee}, which calculated the axial anomaly in an arbitrary even dimensional noncommutative field theory by using the point-splitting regularization. 

The canonical momenta are
\begin{eqnarray}
\Pi_{\psi}&=&i \psi^{\dagger},\\
\Pi^{0}&=&0,  \qquad \textrm{and}\qquad \Pi^{1}=  F^{10} + {3\over2}g\theta (F^{01})^2.
\end{eqnarray}

The equations of motion arising from the Lagrangian~(\ref{schwinger}) are
\begin{equation}
i\Dslash \psi =0 \label{dirac}
\end{equation}
and
\begin{equation}
\partial_{\xi} \left( F^{\lambda \xi} - {3\over 2} g \theta \epsilon^{\alpha\xi}F^{\lambda\delta} F_{\delta\alpha} \right)= g \bar{\psi}\gamma^{\lambda} \psi. \label{eom}
\end{equation}
Eqn.~(\ref{dirac}) is nothing but the Dirac equation. From Eqn.~(\ref{eom}), it it straightforward to show that the current operator $\bar{\psi}\gamma^{\mu} \psi$ is conserved.

\section{Vacuum polarization}
In analogy to commutative gauge theories, perturbative analysis begins with gauge fixing. Feynman rules for the fermion and photon propagators and the fermion-photon vertex are the same as the ones in the commutative Schwinger model. It is straightforward to derive from Eqn.~(\ref{schwinger}) the three-photon vertex and it reads:
\begin{equation}
\begin{split}
V^{\mu\nu\rho}(k,p,q) ={1\over 2} \theta g \epsilon^{\alpha\beta}  \left( k_{\beta} \left( \left(p^{\mu}q^{\nu}-p\cdot q g^{\mu\nu} \right)g^{\rho}_{\alpha} + \left(q^{\mu} p^{\rho} -p\cdot q g^{\mu\rho} \right) g^{\nu}_{\alpha} \right) + p_{\beta} \left( \left( k^{\rho}q^{\nu}-k\cdot q g^{\rho\nu}\right) g^{\mu}_{\alpha} \right. \right. \\
\left. + \left( q^{\mu} k^{\nu}-k\cdot q g^{\mu\nu}\right) g^{\rho}_{\alpha} \right) + q_{\beta} \left( \left( k^{\rho}p^{\mu}-k\cdot p g^{\mu\rho}\right) g^{\nu}_{\alpha} + \left( p^{\rho} k^{\nu}-k\cdot p g^{\nu\rho}\right) g^{\mu}_{\alpha} \right)  \\
\left. -p^{\mu} \left(q_{\alpha}k_{\beta} g^{\nu\rho}+g^{\nu}_{\alpha} g^{\rho}_{\beta} k\cdot q  \right) - p^{\rho} \left(k_{\alpha} q_{\beta} g^{\nu\mu}+g^{\nu}_{\alpha} g^{\mu}_{\beta} k\cdot q  \right)  -q^{\mu} \left(p_{\alpha}k_{\beta} g^{\nu\rho}+g^{\rho}_{\alpha} g^{\nu}_{\beta} k\cdot p  \right) \right. \\
\left. -  q^{\nu} \left(k_{\alpha}p_{\beta} g^{\mu\rho}+g^{\rho}_{\alpha} g^{\mu}_{\beta} k\cdot p  \right) -k^{\nu} \left(q_{\alpha}p_{\beta} g^{\mu\rho}+g^{\mu}_{\alpha} g^{\rho}_{\beta} p\cdot q  \right)-k^{\rho} \left(p_{\alpha}q_{\beta} g^{\mu\nu}+g^{\mu}_{\alpha} g^{\nu}_{\beta} p\cdot q  \right) \right)
\end{split}
\end{equation}
where photon momenta $k^{\mu}$, $p^{\nu}$ and $q^{\rho}$ satisfy $k+p+q=0$.

\begin{figure}[t]
\begin{center}
\includegraphics[width=10cm,clip=true,keepaspectratio=true]{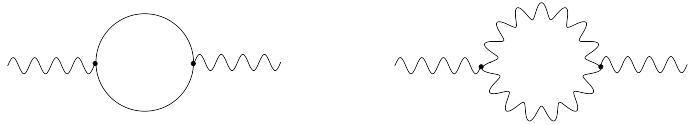}
\caption{\small The lowest-order vacuum polarization diagrams. Wave lines denote photons, solid lines denote fermions.}
\end{center}\label{loop}
\end{figure}    

There are two one-loop diagrams for the self-energy of the gauge boson, as shown in Fig. 1. We do not include the tadpole diagrams since they are identically zero. In the commutative Schwinger model it is the well-known fact that the fermion-loop diagram dynamically generates a mass for the photon field; the result is $m_{\gamma}={g \over \sqrt \pi}$ \cite{schwinger}. The mass generation is due to the IR behavior of the intermediate state formed by a fermion-antifermion pair. Recall that the Lagrangian defined by (\ref{schwinger}) has another Lorentz invariant dimensionful constant, the noncommutativity parameter $\theta$. It is natural at this point to expect that the photon mass is going to be changed by the photon-loop diagram. However, as we shall show by explicit calculation, this is not the case. 

The photon-loop diagram gives
\begin{equation}
{1\over2} \int {d^2p \over (2\pi)^2} V^{\alpha\beta \mu} (-(p+q),p,q ) G_{\alpha\rho}(p+q) V^{\rho\lambda\nu}( p+q, -p, -q ) G_{\beta \lambda}(p)
\end{equation}
where $G_{\alpha\rho}(p+q)$ and $ G_{\beta \lambda}(p)$ are photon propagators and ${1\over2}$ is a symmetry factor.
Note that this integral is gauge independent. By naive power counting, the photon-loop diagram is quadratically divergent. We use the dimensional regularization scheme and evaluate the integrals in $(2-\epsilon)$ dimensional spacetime to extract possible singularities. As a matter of fact, explicit calculation shows that all spurious poles cancel each other and the diagram is well-defined and finite. Remarkably, the degree of divergence is lower than expected because of the gauge symmetry. The same striking phenomenon happens for the fermion-loop diagram, whose superficial degree of divergence is 0. The calculation is straightforward and we report here the final result for the photon-loop diagram:
\begin{equation}  
i \Pi^{\mu\nu}(q) = -i  {(\theta g)^2 \over 2 \pi} q^2 \left(q^{\mu} q^{\nu} - g^{\mu\nu} q^2\right) .\label{loop}
\end{equation}
This result shows that instead of contributing to the photon mass, the radiative correction due to the photon loop generates a new type of operator. The independence of the Schwinger mass on the noncommutativity of spacetime makes sense since the mechanism for the photon mass generation is purely an IR effect. In fact, one can show that the term $i q^2(q^{\mu}q^{\nu}- q^2 g^{\mu\nu})$ corresponds to the dimension four operator $-{1\over 2}\partial_{\mu}F^{\mu\nu}\partial^{\lambda}F_{\lambda\nu}$. Therefore, a higher derivative term is dynamically generated due to the noncommutativity of spacetime. 

Note that at quadratic order in the fields the $\theta$-expanded action of a noncommutative theory is the same as in the commutative theory. This means that the $O(\theta^2)$ part of the classical noncommutative Schwinger model contains only interaction terms. The higher derivative kinetic terms will not show up in the classical action. Thus the appearance of the operator $\partial_{\mu}F^{\mu\nu}\partial^{\lambda}F_{\lambda\nu}$ is a pure quantum effect.

Higher derivative terms are also discovered in Ref.~\cite{bichl}, where the quantization of noncommutative QED via the Seiberg-Witten map is investigated. It is argued that because of the existence of nonrenormalizable vertices in the $\theta$-expanded action, higher derivative terms allowed by symmetries should be added to extend the action in order to absorb the divergences. It is realized later \cite{bichl2} that higher derivative terms are in fact a part of the Seiberg-Witten map. A remarkable difference here is that while the term $\partial F \partial F $ is of order $\theta$ in the noncommutative QED action, it appears in second order in $\theta$ in the noncommutative Schwinger model. This is due to the fact that the coupling constant $g$ is dimensionful in two dimensional spaccetime. Besides, our calculation shows that while the radiative corrections to the photon self-energy are divergent in noncommutative QED, they are finite in the noncommutative Schwinger model. Hence, like the generation of the photon mass, the appearance of the operator $\partial_{\mu}F^{\mu\nu}\partial^{\lambda}F_{\lambda\nu}$ is a dynamical effect.

\section{Unitarity}
We next consider the issue of unitarity. Including the higher derivative term generated by the photon loop into the Lagrangian (\ref{schwinger}), the extended gauge sector becomes\footnote{To avoid the unnecessary complexity, we do not include the term $\sim F_{\mu\nu}{1\over \Box} F^{\mu\nu}$ generated by the fermion loop since the following discussion on unitarity does not depend on the pole of the photon propagator.}
\begin{equation}
- {1\over 4} \left( F_{\alpha\beta} F^{\alpha\beta} + {\sqrt{2 \pi} \over M_{\theta}}  \epsilon^{\alpha \beta}F_{\beta\gamma} F^{\gamma\delta} F_{\delta\alpha} \right) + {1 \over 2 M_{\theta}^{2}}\partial_{\alpha}F^{\alpha\mu}\partial^{\beta}F_{\beta\mu} \label{gauge}
\end{equation}
where $M_{\theta} \equiv {\sqrt{2 \pi} \over g\theta}$. The positive sign of the higher derivative term assures the vacuum stability. $M_{\theta}$ and the photon mass $m_{\gamma}$ are related by the relation
\begin{equation}
M_{\theta} m_{\gamma} ={\sqrt{2} \over  \theta}.
\end{equation}

Theories with the lowest-order higher derivative operators contain states with negative norm, that is, ghost states. In fact, defining \cite{hawking}
\begin{eqnarray}
\tilde{A}_{\mu} &=& {1 \over M_{\theta}^{2}} \left[ \left( \Box + M_{\theta}^{2} \right) A_{\mu} - \partial_{\mu} \left( \partial \cdot A \right) \right],\\
\bar{ A}_{\mu} &=& {1 \over M_{\theta}^{2}} \left[ \Box  A_{\mu} - \partial_{\mu} \left( \partial \cdot A \right) \right],
\end{eqnarray}
Eqn.~(\ref{gauge}) can be rewritten as
\begin{equation}
\begin{split}
 -{1\over 4} \tilde{F}_{\alpha\beta} \tilde{F}^{\alpha\beta} + {1\over 4} \bar{F}_{\alpha\beta} \bar{F}^{\alpha\beta}-{M_{\theta}^{2} \over 2} \bar{A}_{\alpha}\bar{A}^{\alpha}
-\sqrt{{ \pi\over 8}}{1\over M_{\theta}} \epsilon^{\alpha\beta} & \left( \tilde{F}_{\beta\gamma} \tilde{F}^{\gamma\delta} \tilde{F}_{\delta\alpha}- \bar{F}_{\beta\gamma} \bar{F}^{\gamma\delta} \bar{F}_{\delta\alpha} \right. \\ 
& \left. -3  \tilde{F}_{\beta\gamma} \tilde{F}^{\gamma\delta} \bar{F}_{\delta\alpha} +3 \bar{F}_{\beta\gamma} \bar{F}^{\gamma\delta} \tilde{F}_{\delta\alpha}\right). \label{gauge2}
\end{split}
\end{equation}
The unusual positive sign of the kinetic term of the field $\bar{A}_{\mu}$ means that $\bar{A}_{\mu}$ particles are massive ghosts. The last two terms in Eqn.~(\ref{gauge2}) connect the two Hilbert spaces where $\tilde{A}_{\mu}$ and $\bar{A}_{\mu}$ particles live respectively and cause the nonconservation of the ghost number. This would result in the loss of unitarity. Note that different from the photon mass, which is independent of the noncommutativity parameter $\theta$, the mass of ghost fields $\bar{A}_{\mu}$ depends on both dimensionful parameters in the theory. Since states with $\bar{A}_{\mu}$ particles decouple from the theory in the limit $\theta \rightarrow 0$, unitarity violation is caused by noncommutativity of spacetime. This is consistent with the well-known fact that $\theta^{0i} \neq 0$ leads to unitarity violation \cite{gomis}.

However, as was argued in Ref.~\cite{hawking}, a physical $S$ matrix defined between stable states is still well-behaved as long as ghost particles can decay and do not show up as asymptotic states. In the noncommutative Schwinger model this requires
\begin{equation}
m_{\gamma}^{2}\theta \leq {1\over \sqrt{2}}.
\end{equation}
As a result, an interacting field theory with higher derivative operators can still make sense even though unitary is lost at a high energy scale of order $M_{\theta}$, the ghost mass. In that sense, even though interaction terms between massive ghost fields and physical fields cause a flaw in the theory, they provide a cure in the meantime.

As a final point, let us observe the higher loop effects. Because of the existence of the three-photon interaction term, which is non-renormalizable by naive power counting, higher-order 1PI vacuum polarization diagrams will in principle generate an infinite number of higher derivative operators. However, it is not a difficult exercise to show that a theory with higher derivative kinetic terms can always be rewritten as another equivalent theory which involves more fields (at least one of them is a ghost if the lowest-order higher derivative kinetic terms exist in the original theory) and contains terms with at most two derivatives in it. Therefore, the above argument still holds.

Even though we have only done calculations for the noncommutative Schwinger model, we have reasons to believe the emergence of higher derivative operators is a feature for other theories with space-time noncommutativity. This is because for any theory expansions in $\theta$ absolutely contain nonrenormalizable vertices. As a result, by power-counting analysis higher derivative operators obeying symmetries will be generated in loop diagrams. Since higher derivative kinetic terms obey the same symmetries as the classical two-derivative kinetic terms, they shall naturally be generated by the polarization diagrams.

Thus, from the above discussion we conclude that in the framework of the Seiberg-Witten map the feature of non-unitarity for a noncommutative field theory with $\theta^{0i}\neq 0$ is characterized by the presence of higher derivative kinetic terms. That is, the $\theta$-expanded version of a unitary theory will not generate the lowest-order higher derivative kinetic terms.

\section{Conclusion}
The purpose of this paper has been to show via a study of the noncommutative Schwinger model that the noncommutativity of space-time, though still waiting for the experimental proof, provides an explanation for the emergence of higher derivative kinetic terms in a field theory. Our focus on the noncommutative Schwinger model bypasses unrelevant complexities as can be seen by comparing Eqn.~(\ref{1staction}) and Eqn.~(\ref{schwinger}). This simplicity is mainly because a two dimensional fermion is insensitive to the noncommutativity of spacetime up to order $O(\theta)$. In addition, issues related to the breakdown of Lorentz invariance are avoided in two dimensional spacetime.

As an aside, we showed that the Schwinger mass is not modified by the noncommutativity of spacetime till the first nontrivial order in $\theta$.

\begin{acknowledgments}
 The research of F.W. was supported in part by the National Nature Science Foundation of China under grant No. 10805024 and the project of Chinese Ministry of Education under grant No. 208072. M.Z. was supported in part by the research fund of National University of Defense Technology.
\end{acknowledgments}


\end{document}